\documentclass[sigconf,authordraft]{acmart}
\renewcommand\footnotetextcopyrightpermission[1]{} 
\usepackage{caption}
\usepackage{subcaption}
\AtBeginDocument{%
  \providecommand\BibTeX{{%
    \normalfont B\kern-0.5em{\scshape i\kern-0.25em b}\kern-0.8em\TeX}}}

\setcopyright{acmcopyright}
\copyrightyear{2020}
\acmYear{2020}
\acmDOI{10.1145/1122445.1122456}

\acmConference[CIKM '20]{CIKM '20: 29TH ACM INTERNATIONAL CONFERENCE ON INFORMATION AND KNOWLEDGE MANAGEMENT}{Oct 19--23, 2020}{Galway, Ireland}
\acmPrice{15.00}
\acmISBN{978-1-4503-9999-9/18/06}

\usepackage{bm}
\usepackage{amsfonts}
\usepackage{multirow}
\usepackage{xcolor}
\usepackage{soul}

\begin{document}

\title{SERank: Optimize Sequencewise Learning to Rank Using Squeeze-and-Excitation Network}

\author{Ruixin Wang}
\affiliation{Zhihu Search}
\email{wangruixin@zhihu.com}

\author{Kuan Fang}
\affiliation{Zhihu Search}
\email{fangkuan@zhihu.com}

\author{Rikang Zhou}
\affiliation{Zhihu Search}
\email{zhourikang@zhihu.com}

\author{Zhan Shen}
\affiliation{Zhihu Search}
\email{shenzhan@zhihu.com}

\author{Liwen Fan}
\authornote{The work was done when the author was with Zhihu Search}
\email{levyfan@163.com}


\begin{abstract}
Learning-to-rank (LTR) is a set of supervised machine learning algorithms that aim at generating optimal ranking order over a list of items. A lot of ranking models have been studied during the past decades. And most of them treat each query document pair independently during training and inference. Recently, there are a few methods have been proposed which focused on mining information across ranking candidates list for further improvements, such as learning $multivariant$ scoring function or learning contextual embedding. However, these methods usually greatly increase computational cost during online inference, especially when with large candidates size in real-world web search systems. What's more, there are few studies that focus on novel design of model  structure for leveraging information across ranking candidates. In this work, we propose an effective and efficient method named as SERank which is a Sequencewise Ranking model by using Squeeze-and-Excitation network to take advantage of cross-document information. Moreover, we examine our proposed methods on several public benchmark datasets, as well as click logs collected from a commercial Question Answering search engine, Zhihu. In addition, we also conduct online A/B testing at Zhihu search engine to further verify the proposed approach. Results on both offline datasets and online A/B testing demonstrate that our method contributes to a significant improvement.
\end{abstract}


\maketitle
\pagestyle{plain} 

\section{Introduction}
In the past decades, a plenty of learning to rank (LTR) algorithms have been studied and applied in search engine systems, where the task of these methods is to provide a score for each document in a list for a given query, so that the documents ranked higher in the list are expected to have higher relevance. The majority of ranking methods take each document's feature as input and learn a scoring function by optimizing loss functions which could be categorized into $pointwise$ \cite{chen2009ranking}, $pairwise$ \cite{burges2010ranknet} and $listwise$ \cite{xia2008listwise}. In the last few years, benefiting from the powerful nonlinear representation ability of deep learning methods \cite{csaji2001approximation} as well as the huge amount of web data, deep ranking models have been widely proposed and deployed in many real-world scenarios \cite{haldar2019applyingairbnb}.

\begin{figure}[h]
    \centering
    \includegraphics[scale=0.25]{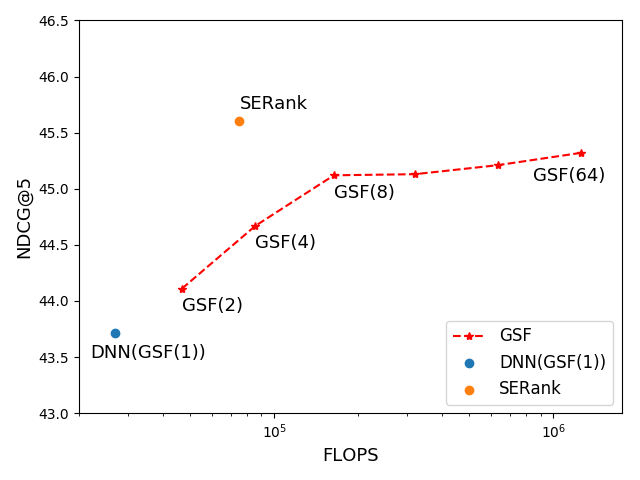}
    \caption{\small{FLOPS vs NDCG@5 between different models on Web30K dataset. We use GSF(.) to represent the Groupwise Ranking model with different group size, and DNN(GSF(1)) model is the standard feed forward neural network with three fully-connected layers. The SERank model outperforms all of the GSF models on metric NDCG@5 with a speed of 16.7x faster than the GSF(64) model.}}
    \label{fig:flops}
\end{figure}

\begin{figure}[h]
    \begin{subfigure}[b]{0.45\textwidth}
    \centering
    \includegraphics[width=\textwidth]{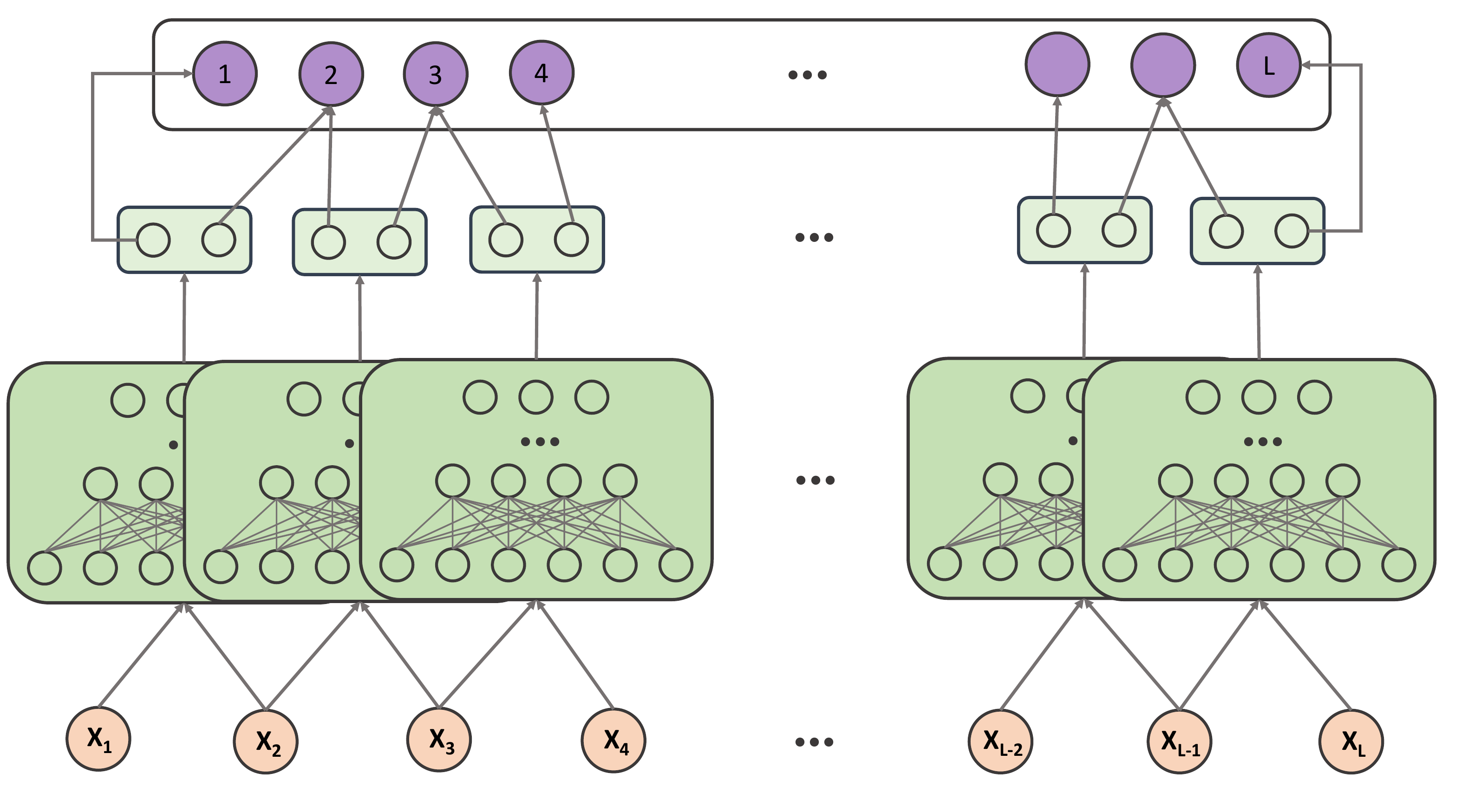}
    \caption{\small{GSF(2), groupwise model with group size = 2}}
    \label{fig:groupwise}
    \end{subfigure} 
    \begin{subfigure}[t]{0.23\textwidth}
    \centering
    \includegraphics[width=\textwidth]{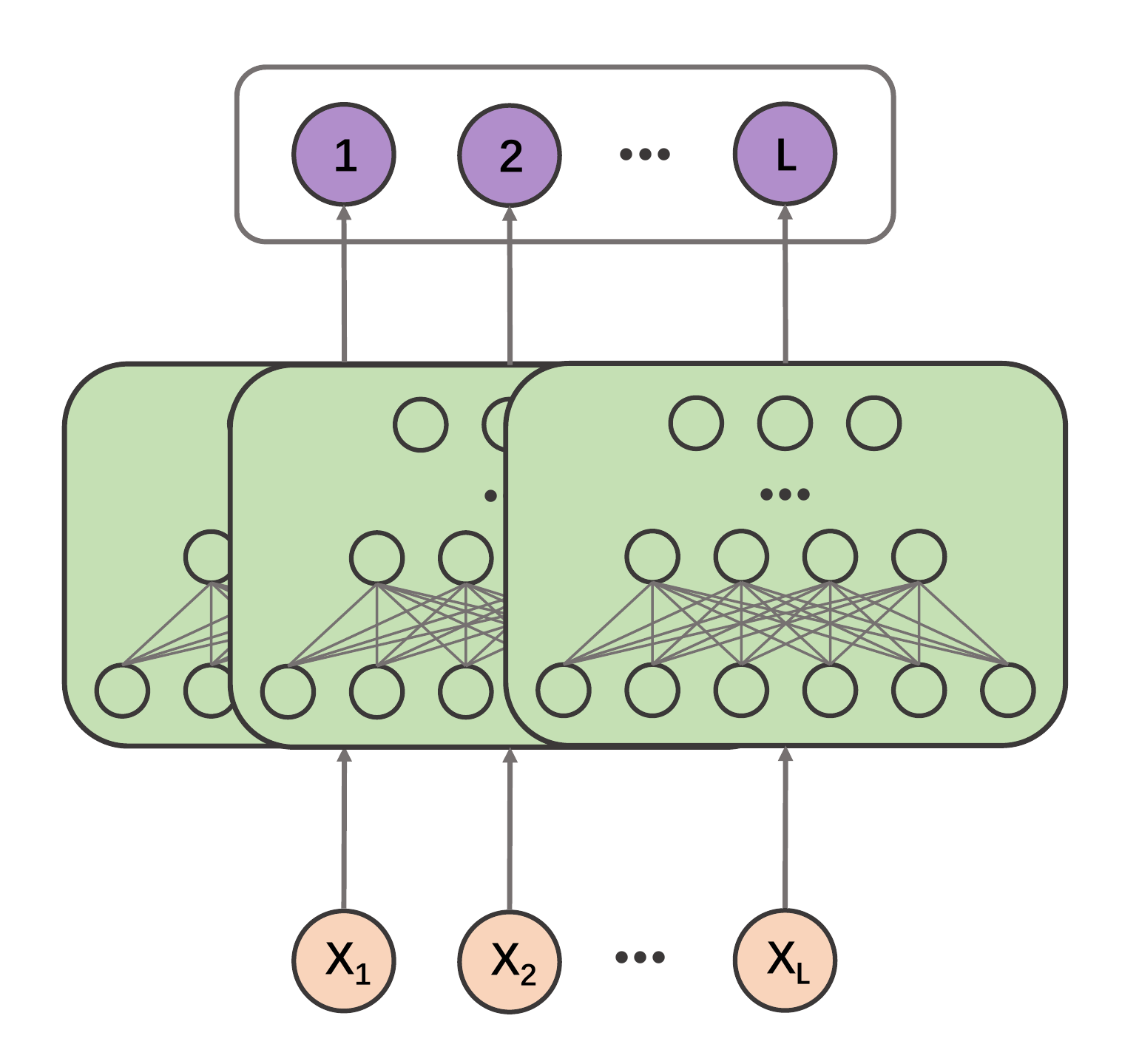}
    \caption{\small{basic DNN model, equivalent to GSF(1)}}
    \label{fig:baseline}
    \end{subfigure}
    \begin{subfigure}[t]{0.165\textwidth}
    \centering
    \includegraphics[width=\textwidth]{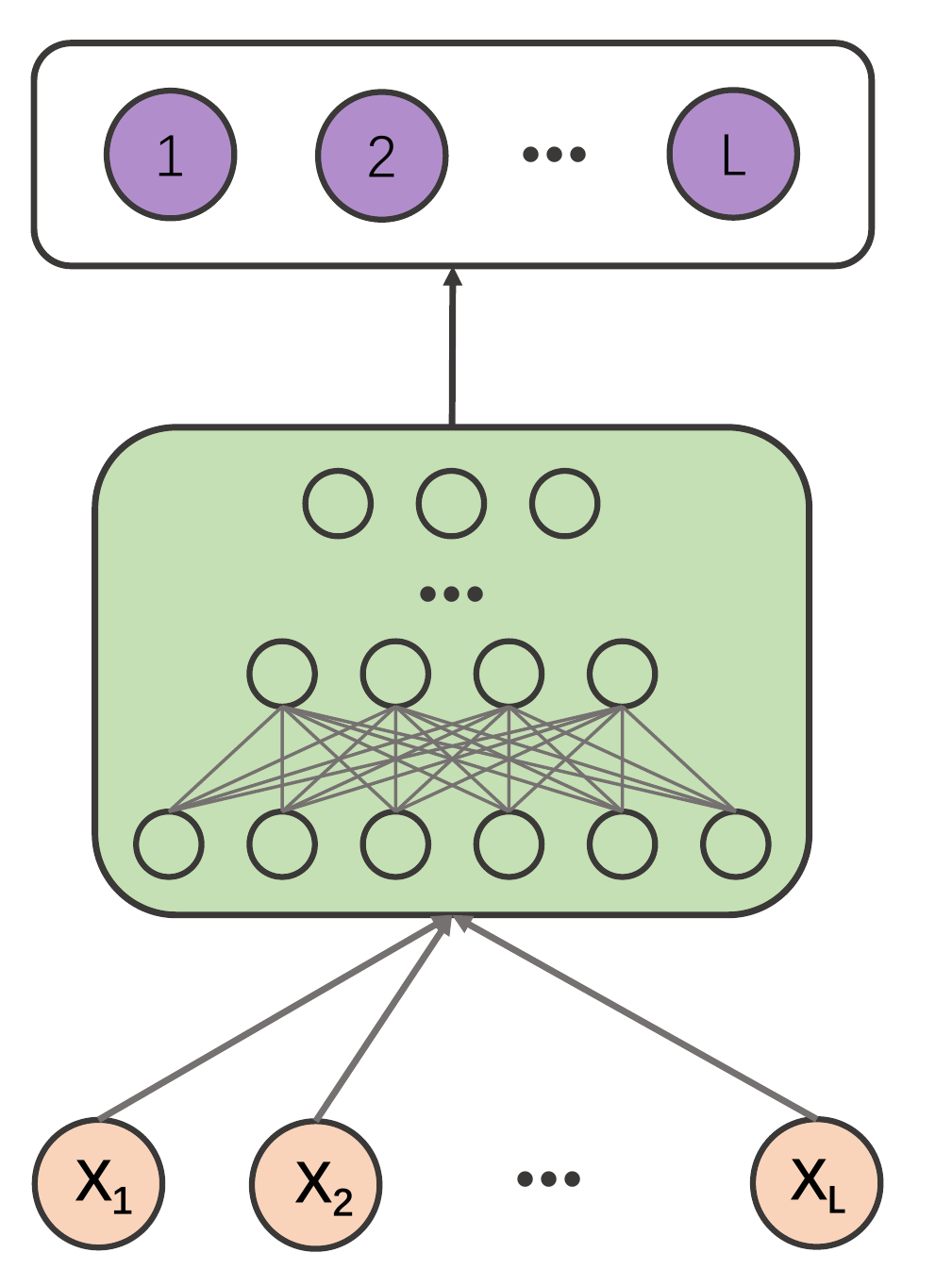}
    \caption{\small{Sequencewise model}}
    \label{fig:sequencewise}
    \end{subfigure}
\caption{Score mechanisms of Groupwise Ranking (a), basic DNN model (b)  and Sequencewise Ranking (c). The Groupwise model accepts input features grouped by documents and output a group of relevance scores, then it aggregates each item's score among different groups to get the final score. The basic DNN model accepts each document's input feature and outputs single score corresponding to it. The proposed Sequencewise model takes a sequence of documents as input, and jointly outputs their final scores.}
\end{figure}

\indent Recently, leveraging information across ranking documents becomes an emerging topic in the LTR domain. A number of works have proven that cross-document information could be mined to enhance final ranking performance \cite{ai2018learning}  \cite{ai2019learning} \cite{bello2018seq2slate}. Qingyao et al. \cite{ai2018learning} defines the $multi\-variant$ scoring functions which named as GSF (Groupwise Scoring Function) by feeding concatenated features among a group into the DNN model, so that information across documents could be automatically learned. However, the GSF ranking with large group size increases model complexity and results in an huge expansion of computation cost, which makes it unappealing for real-world online services with sensitive responding time, while small group size often results in an insignificant gain of ranking quality. \newline
\indent  In this paper, to tackle the problem of utilizing cross-document information efficiently, we define a new Sequencewise ranking model named as SERank\footnote{Our source code will be released soon.} which jointly scores and sorts a sequence of ranking candidates at once. As shown in Figure \ref{fig:sequencewise}, the proposed SERank takes a sequence of documents as input and scores them jointly, rather than predicts each document individually. Furthermore, feature importance, which is crucial for LTR settings, may vary when the feature distribution of ranking candidates varies. Therefore, Squeeze-and-Excitation \cite{hu2018squeeze} is introduced to learn the feature importance information dynamically from a sequence of ranking candidates in each query. Figure \ref{fig:flops} summarizes the efficiency and effectiveness of SERank with other models. Our SERank model surpasses all the baseline models on the NDCG metric \cite{jarvelin2017ir}. On the efficiency aspect, the SERank runs 16.7x faster than GSF(64) model. Finally, different from the architecture proposed in \cite{ai2018learning}, our proposed method does not require an initial ranking order over the candidates. The SERank model could accept any arbitrary permutations of input document sequence and jointly output score for each document.\newline
\indent In summary, our main contributions are listed as follows:

\begin{itemize}
    \item We define a Sequencewise ranking model called SERank, which scores all ranking candidates in one episode and does not require an initial ranking order over the candidate documents. 
    \item We introduce Squeeze-and-Excitation Network into the LTR settings for mining feature importance information across ranking candidates and propose the concrete implementation of the SERank model.
    \item Results on different benchmark datasets and online A/B testing illustrate our designed model obtains better ranking quality and requires little additional computations cost. 
\end{itemize}

\indent The rest of this paper is organized as follows. In Section \ref{related_work}, we review related works that are relevant to our proposed model. After that, we formally define the research problem and explain our proposed SERank model in Section \ref{problem} and Section \ref{model}. We will present experimental explorations on offline benchmark datasets as well as online A/B testing in Section \ref{experiment}. Finally, we discuss empirical results and conclude this work in Section \ref{conclusion}.

\section{Related Work} \label{related_work}
Overall speaking, there are four main subjects of research that are related to the work in this paper: they are studies on learning to rank, nerual ranking models, nerual re-ranking models and Squeeze-and-Excitation Network. 

\subsection{Learning to Rank}
Learning to rank \cite{liu2009learningtorank} refers to methods that provide an order for a list of ranking candidates via machine learning approaches. Most research in LTR fields could be categorized by two aspects: the structure of scoring function and the type of loss function. Scoring functions can be parameterized by Gradient Boosting Trees \cite{ke2017lightgbm}, Support Vector Machine \cite{joachims2006training}  \cite{joachims2017unbiased}, and Neural Networks \cite{burges2010ranknet}. Airbnb successfully deployed a feed-forward neural network as a replacement of decision trees in their search system \cite{haldar2019applyingairbnb}. While loss functions can be generally categorized into three types as $pointwise$ \cite{chen2009ranking}, $pairwise$  \cite{burges2010ranknet} and $listwise$ \cite{xia2008listwise}. The $pointwise$ loss function treats LTR as a classification problem where each relevance grade corresponds to one class label. $Pairwise$ loss functions learn document preference between each document pair in a query, and different weight are assigned to each pair according to their relevance label and ranking positions. $Listwise$ methods put together all documents in a query and optimize ranking metrics directly. \newline 
\indent Recently, Ai et al.\cite{ai2018learning} proposed a multi-variant scoring function that scores query documents by a learned $Groupwise$ function, which takes cross-document interatcions into account.
While the name $listwise$ is a type of loss function, and $Groupwise$ means a type of scoring function here,  our Sequencewise model focuses on the improvement on the model structure.   \newline

\subsection{Neural Ranking Models}

In learning to rank circle, LambdaMART has been the long-standing state-of-the-art model for past decades. However, with the tremendous growing amount of data on the web, building a more effective ranking model with millions or tens of millions of training data becomes one challenging problem. \newline
\indent The neural network based models have the capability of learning from large scale data and high flexibility to the type of input features. For instance, in the scenario of Airbnb search \cite{haldar2019applyingairbnb}, a neural network with single hidden layer and 32 fully connected ReLU activations obtains comparable result against tree based ranking models, and a deep NN model with 2 hidden layers and 10x larger training data gains significant improvement over GBDT. Moreover, the NN based model has the advantage of learning representations of sparse id features. In the scenario of Gmail Search \cite{pasumarthi2019tfranking}, they achieve substantial improvement over the baseline model after adding sparse token id features of each query and document, while the tree based model often fails to encode this kind of feature type. Therefore, in the condition of large scale training data and various kinds of feature types, the neural network based models have been proven as superior models against the traditional LambdaMART model. \newline 
\indent Despite the fact that the NN model has prove to surpass GBDT in the scenarios described above, there are a lot of efforts\cite{ke2019deepgbm} \cite{li2019combining} \cite{ling2017model} focused on combining GBDT and NN models together for further enhancement. These methods are designed to distil advantages of NN and GBDT respectively and then makes the combined model more powerful than each separate model. However, in this paper we focus on designing new model structure on NN aspect, and this model structure could be easily incorporated into a GBDT-NN combined framework.

\subsection{Neural Re-Ranking Models}
Recently, different from the traditional global ranking model that scores the entire ranking candidates purely based on each candidate's own features, a few works\cite{ai2018learning}\cite{ai2019learning}\cite{bello2018seq2slate}\cite{pasumarthi2019attentive}\cite{pei2019prm} have established that the final ranking performance of the top results could be further refined by utilizing a re-ranking stage based on the ranking order of a global ranking model. Most of the re-ranking methods focus on designing cross-document interactions which makes model captures more information about the whole ranking list. For instance, Ai et al. \cite{ai2018learning} designed a $listwise$ re-ranking algorithm by using RNN to extract additional context-aware features for top candidates of a ranked list . Bello et al. \cite{bello2018seq2slate} treats the re-ranking model through a sequence-to-sequence model so that the ranking order of the top candidates are given by the generation order of the decoder in the sequence-to-sequence model. \newline 
\indent However, these methods require a strong initial ranking order of the input candidates (for example, utilizing LambdaMART to select top tens of documents before re-ranking in \cite{ai2019learning}), which introduce additional computation overhead, while our proposed method is served as a global ranking method which scores the entire ranking candidates and could be combined with other arbitrary re-ranking methods. 


\subsection{Squeeze-and-Excitation Network}
In the computer vision circle, the Squeeze-and-Excitation network \cite{hu2018squeeze} has been proposed and applied successfully, where the main intuition is to learn inter-dependencies between different feature channels for an image. This kind of mechanism contributed a winning of first prize in the ILSVRC 2017 \cite{park2017ilsvrc} classification task.\newline
\indent There are a few variants of the original Squeeze-and-Excitation network \cite{li2019selective}\cite{cao2019gcnet}. For example, Li et al. propose SKNet \cite{li2019selective} which collect local information with multi branches of different receptive field size, and merge different branches by the learned attention weights. With a little increase in model complexity, this method outperforms many state-of-art architectures.  \newline 
\indent Recently, Huang et al. \cite{huang2019fibinet} successfully incorporate the Squeeze-and-Excitation network in the recommendation ranking model. However, they use SENet to learn feature importance weights only by each recommendation item's feature embedding, while we focus on using SE block to design a Sequencewise ranking model.
\begin{figure}
    \includegraphics[scale=0.5]{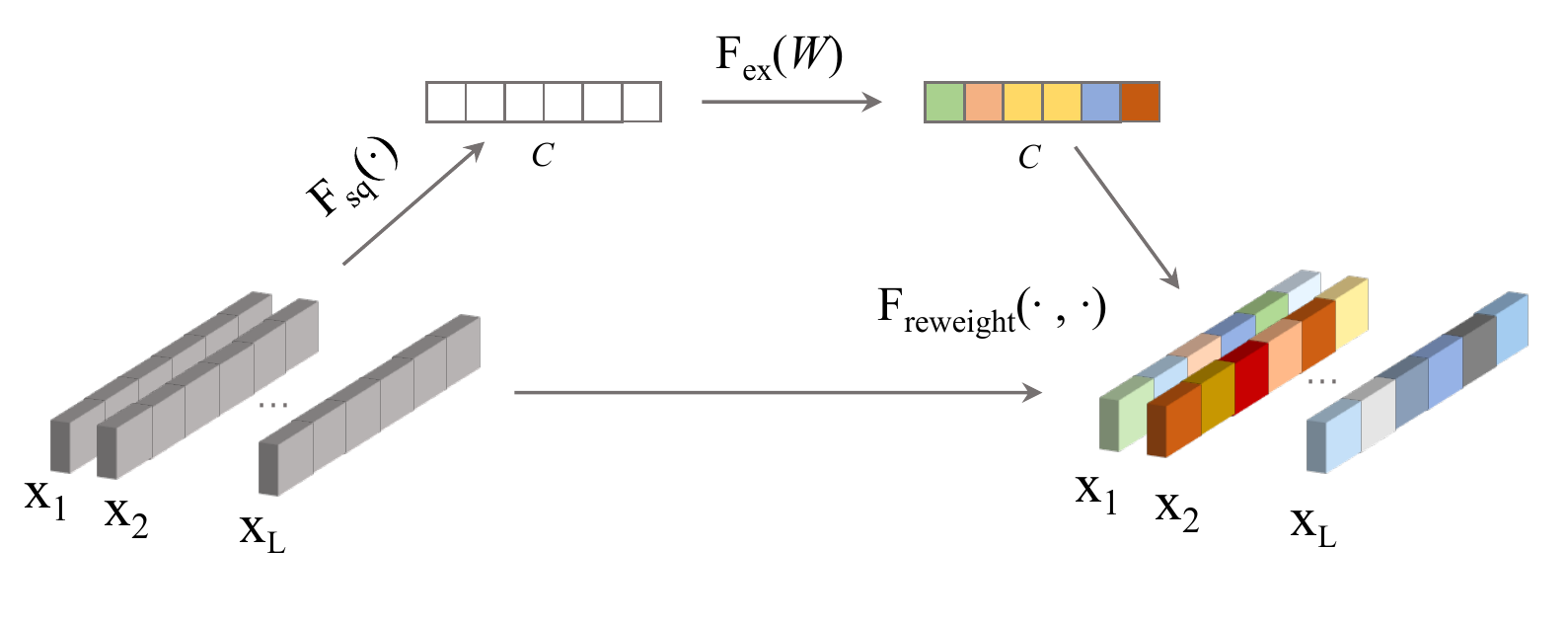}
    \caption{\small{A Squeeze-and-Excitation block. $\{x_1, x_2,...,x_L\}$ is the input documents for a given query. The squeeze operation $F_{sq}(.) $ gathers information across feature channels among documents and the excitation operation $F_{ex}(W)$ gain the feature importance. Finally, the raw inputs $x_i$ are re-weighted by the output of $F_{ex}(W)$.}}
    \label{fig:seblock}
\end{figure}

\section{Problem Formulation} \label{problem}
In the learning-to-rank settings, the training set could be represented as $\Psi=\{\bm{x},\bm{y}\} \in \textit{X}^n \times \mathbb{R}^n$, where $\bm{x}$ is a vector of $n$ items $\bm{x}_i$, $1 \leq i \leq n$, $\bm{y}$ is also a vector of $n$ real values that represents the relevance grade of each $\bm{x}_i$, and $\textit{X}$ is the space of all items. In this paper, we denote $\bm{x}_i$ as a vector of features represents for a query document pair. The main goal of a ranking model is to learn a scoring function, which maps the input feature vector of $\bm{x}_i$ to a real output value $R_i$, and the score function should minimize the empirical loss over the training set, which could be formally stated via a supervised machine learning framework
\begin{equation}
\mathcal{L}(f)= \frac{1}{\Psi}\sum_{\bm{x},\bm{y}\in \Psi}\mathnormal{l}\{\bm{y},f(\bm{x})\},
\end{equation}
where $l(.)$ is a loss function over training examples, and $f(.)$ is usually an $univariate$ function which accepts single item $\bm{x}_i$ as input.  \newline
\indent Most research focuses on optimizing loss functions or model structures that still rank each document independently. Recently a few works contributed on how to leverage cross-document information to further improve the ranking performance. The ranking model predicts document relevance with additional information either by using $multivariate$ \cite{ai2019learning} score function which receives aggregated items $\{\bm{x}_i,\bm{x}_j...\bm{x}_k\}$ as inputs or learning additional context features for each item $\bm{x}_i$ \cite{ai2019learning}. Besides these approaches, we propose SERank to leverage cross-document information. Generally, our methods could be represented as a learning process as follows
\begin{equation}
\mathcal{L}(F)= \frac{1}{\Psi}\sum_{\bm{x},\bm{y}\in \Psi}\mathnormal{l}\{\bm{y},F(\bm{x}, g(\sum_{\bm{x_i} \in \Psi_q}\bm{x_i}))\} .
\end{equation}
We define a function $g(\sum_{\bm{x_i} \in \Psi_q}\bm{x_i})$ to capture feature importance through aggregating feature info across documents in each query, where $\Psi_q$ is the documents corresponding to the given query $q$. Then result from $g(\sum_{\bm{x_i} \in \Psi_q}\bm{x_i})$ and raw input $\bm{x}$ are combined together to the final score function $F(.)$. We will describe the details in the next sections.

\section{Sequencewise Deep Ranking Model} \label{model}
In this section, we describe how our proposed Sequencewise model works in the learning-to-rank regime. Our main intuition is to let model learn feature importance automatically by gathering feature information across ranking documents in each query. Then we incorporate learned feature importance information into the ranking model. 

\subsection{Sequencewise Input Layer}
Before describing model structure, we first explain the details of the input layer. In the LTR settings, the documents are grouped by queries, and for each query, the data can be organized as a set of documents $\{\bm{x}_{(1)}, \bm{x}_{(2)}, ..., \bm{x}_{(L)}\}$ corresponding to it. Each document $\bm{x}_i$ could be represented as a feature vector $\{\bm{x}^1_{(i)}, \bm{x}^2_{(i)},..., \bm{x}^c_{(i)}\}$ , where $\bm{x}_{(i)}^c$ is the $c$-th feature channel of document $\bm{x}_{(i)}$. Therefore, the ranking model accepts documents $\mathcal{D} \in \mathbb{R}^{L*C}$ and output scores $\mathcal{S} \in \mathbb{R}^{L}$ for all documents per query. In this paper, we use $\bm{C}$ to denote the size of feature channels of each query-document pair, and $\bm{L}$ to denote number of ranking candidates for each query. 

\subsection{The SE (Squeeze-and-Excitation) Block}
According to the motivation proposed in GSF\cite{ai2019learning}, the relevance of each document
depends on the distribution of the whole list. For instance, consider an ad-hoc document retrieval scenario where a user is searching for the name of an artist. If all the results returned by the query (e.g., “calvin harris”) are recent,
the user may be interested in the latest news or tour information. If, on the other hand, most of the query results are older (e.g., “frank sinatra”), it is more likely that the user seeks information on artist discography or biography. So in the ranking tasks, feature importance may vary when met with different candidate lists. Therefore, the main intuition of our method is to design a supplementary block for learning feature importance across ranking documents. \newline
\indent Inspired by SENet \cite{hu2018squeeze}, we apply SE block in our proposed model for feature importance learning. As shown in Figure \ref{fig:seblock}, the SE block computes feature weight through two main mechanisms, gather Sequencewise information by squeeze operation and gain feature importance by excitation operation. 

\subsubsection{Squeeze Operation}
The squeeze operation is designed to compute each feature's statistics over different documents for a given query. To be specific, the squeeze operation process input $\mathbb{X} \in \mathbb{R}^{L*C}$ and then output collected feature statistics $\bm{U} \in \mathbb{R}^{C}$. The squeeze operation could be either max pooling or mean pooling over each feature among different documents.

\subsubsection{Excitation Operation}
After aggregating feature information over documents, the excitation operation aims to generate each feature's importance weight. Two fully-connected layers are used to learn the feature weights. In the first layer, we reduce the dimension with a shrinkage parameter $r$ so that the output of first fully-connected layer is $\mathbb{R}^{\frac{C}{r}}$. Then we recover the dimension with the same $r$ in the next fully-connected layer. Concretely, the final feature weight could be calculated as below
\begin{equation}
    \bm{s} = \bm{F}_{ex}(\bm{U}) = \sigma{1}(\bm{W_2}\times\sigma{2}(\bm{W_1}\times\bm{U})),
\end{equation}
where $\bm{s} \in \mathbb{R}^{\bm{C}}$ is the learned feature weights, $\sigma{1}$ and $\sigma{2}$ are ReLu activation and $\bm{W_1}$ and $\bm{W_2}$ are parameters of two fully-connected layers. After the excitation operation, the SE block use the $\bm{s}$ to re-weight the input layers, by element-wise multiplying the raw input $\mathbb{X} \in \mathbb{R}^{L*C}$ by the excitation output $\bm{s} \in \mathbb{R}^{\bm{C}}$. \newline 
\indent Finally, after the Squeeze and Excitation operations, the important feature channels are strengthened while uninformative feature channels are decreased.

\subsection{Ranking with Squeeze-and-Excitation Block}
\begin{figure}
    \centering
    \begin{subfigure}[b]{0.225\textwidth}
        \centering
        \includegraphics[width=\textwidth]{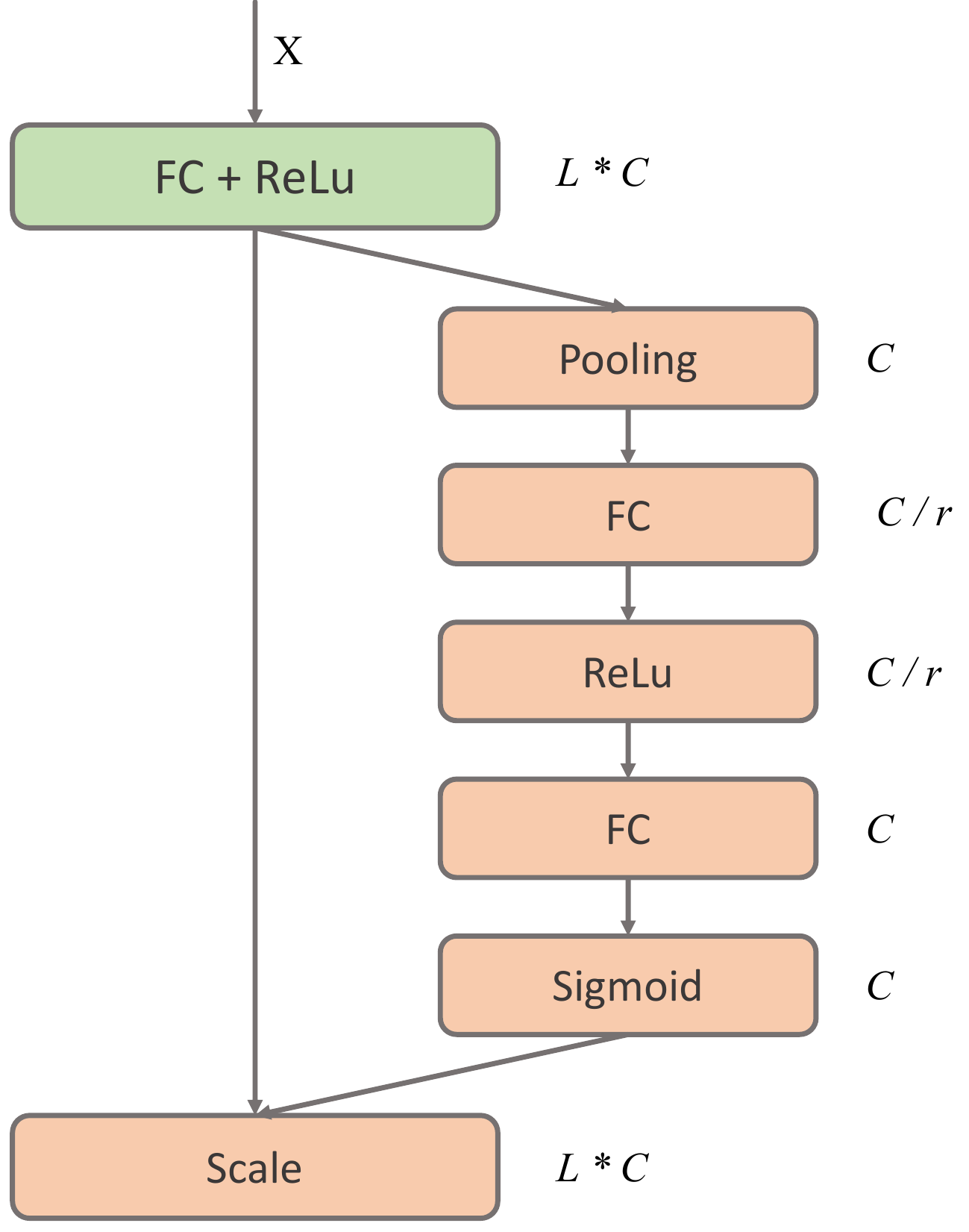}
        \caption{\small{SE block structure}}
        \label{fig:senet-a}
    \end{subfigure}
    \begin{subfigure}[b]{0.237\textwidth}
        \centering
        \includegraphics[width=\textwidth]{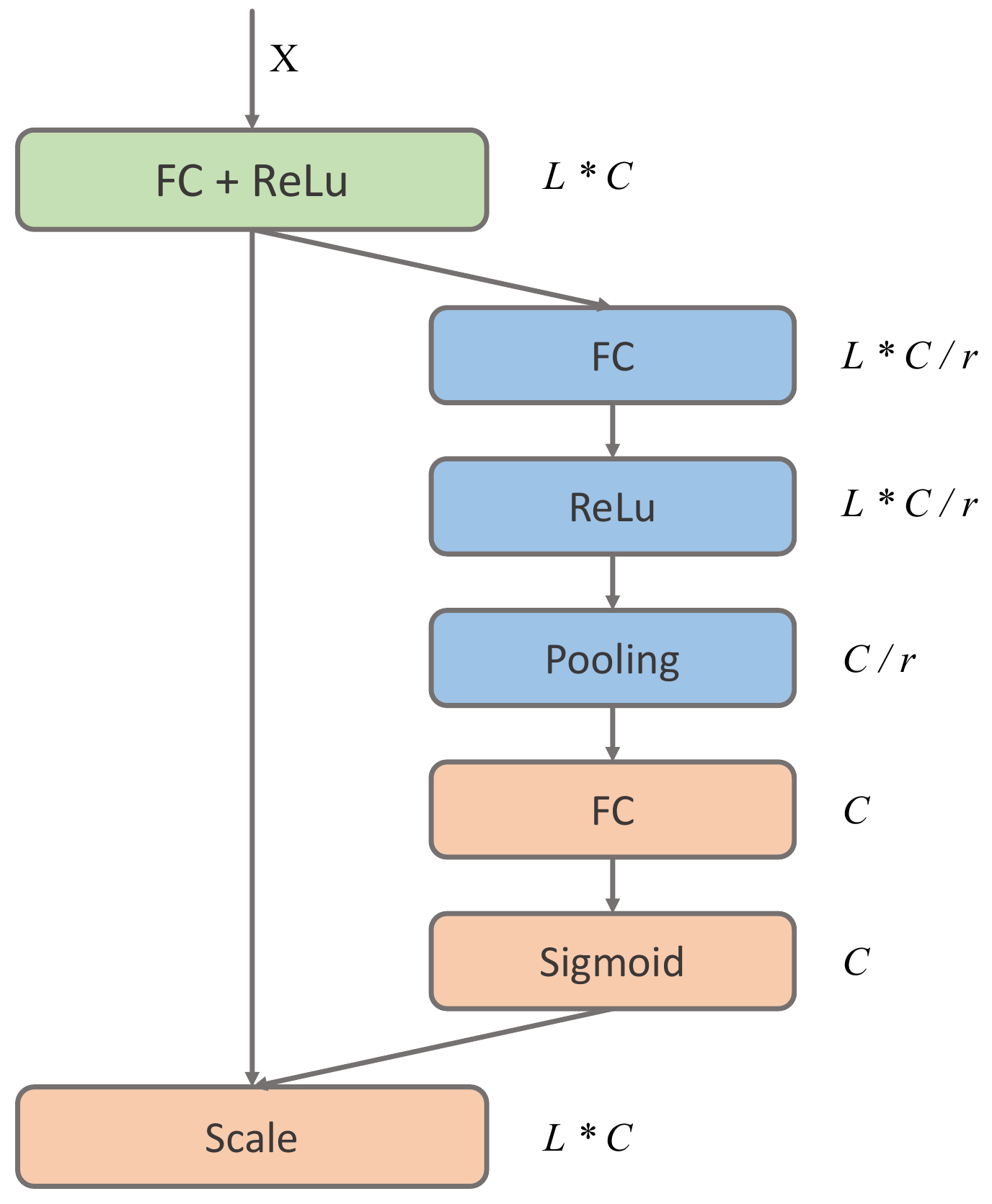}
        \caption{\small{SE-b block structure}}
        \label{fig:senet-b}
    \end{subfigure}
    \caption{The original SE block structure (a) and the modified SE block structure (b)}
\end{figure}
\begin{figure}
    \centering
    \includegraphics[scale=0.28]{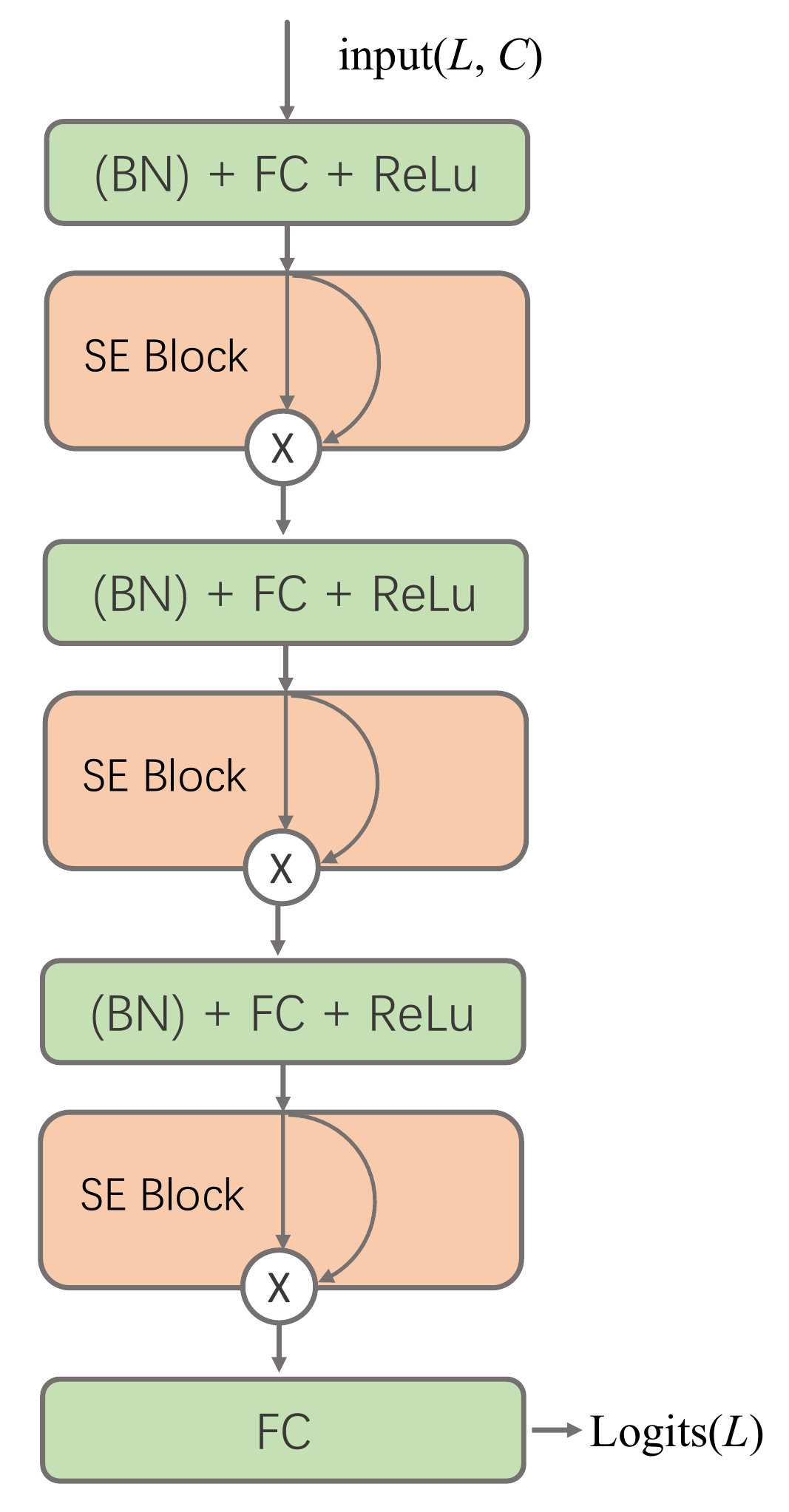}
    \caption{\small{Overall SERank Model Structure. The BN is optional which is only used in Web30K and Web10K dataset.}}
    \label{fig:network}
\end{figure}

\indent Primarily, we use a multi-layer DNN model as the basic model structure for ranking, where all layers are fully-connected layer with different hidden units size. We incorporate SENet into our model by adding the SE block in each layer. The model structure is shown in Figure \ref{fig:network}. \newline
\indent In this paper, different from the original SE block proposed in \cite{hu2018squeeze}, we design a modified SE block version denoted as SE-b. As shown in Figure \ref{fig:senet-b}, SE-b first uses a fully-connected layer to reduce the dimension of each query-document pair, after that, the pooling is adopted for feature information gathering. The reason why we try this minor modification is to let the model gather global feature weight from a compressed input space, rather than from raw input of the last hidden layer directly. We conduct an experiment in the next section and the result shows that our modification leads to further improvement.

\subsection{Loss Functions}
Our proposed model could be trained through any arbitrary loss functions. To verify the generality of SERank, we train with both $pairwise$ and $listwise$ loss functions. Since in LTR the documents are grouped with queries, the loss functions listed here are described for query document pairs within a query.\newline
\indent The first one is the Pairwise Logistic Loss \cite{burges2010ranknet} which is one of the most classic loss functions in LTR field
\begin{equation}
    \mathcal{L}(\bm{y};\bm{\hat{y}}) = \sum_{i=1}^{n-1}\sum_{j=1,\hat{y}_j<\hat{y}_s}^{n}log(1 + exp^{-(\hat{y}_i - \hat{y}_j)}),
    \label{equation:pairwise-loss}
\end{equation}
where the subscript $i$ represents the $i$-th document in a query, $y_i$ is the true label of document $i$ and $\hat{y}_i$ is the predicted score. Note that the Pairwise Logistic Loss can be extended to a $listwise$ loss function by multiplying $\lambda$-weight \cite{wang2018lambdaloss} on the document pairs. Therefore, we also examined Pairwise Logistic Loss with $\lambda$-weight in the experimental section.\newline 
\indent The second loss function is the Softmax Cross-Entropy Loss \cite{bruch2019softmaxloss}, which is a $listwise$ loss calculated as
\begin{equation}
        \mathcal{L}({\bm{y};\hat{\bm{y}}}) =  \sum_{i=1}^{n}\frac{y_i}{\sum_i{y_i}}log(\frac{exp(\hat{y}_i)}{{\sum_i{exp(\hat{y}_i)}}}).
\end{equation}

\begin{table*}[h!]
  \caption{\small{Comparison of test NDCG with baseline models on Web30K and Web10K dataset. For GSF and SERank models, we use softmax loss function to train the model.}}
  \small
  \label{tab:web30k}
  \begin{tabular}[c]{l|l|ccc}
    \toprule
    Dataset & Model & NDCG@1 & NDCG@5 & NDCG@10 \\
    \midrule
    \multirow{8}*{Web30K} & LambdaMART (LightGBM, best reported \cite{bruch2019revisiting}) & \textbf{50.33} & \textbf{49.2} & \textbf{51.05} \\
     & RankNet & 40.74($\pm{0.11}$) & 42.1($\pm{0.05}$) & 44.73($\pm{0.06}$)\\
     & GSF(1) (best reported \cite{ai2019learning}) & - & 43.14 & -\\
     & GSF(1) (fine-tuned) & 43.33($\pm{0.13}$) & 43.70($\pm{0.07}$) & 46.08($\pm{0.05}$)\\
     & GSF(64) (best reported \cite{ai2019learning}) & 44.21 & 44.46 & 46.77\\
     & GSF(64) (fine-tuned) & 45.01($\pm{0.09}$) & 45.32($\pm{0.10}$) & 47.67($\pm{0.08}$)\\
    \cline{2-5}
     & SERank & 44.38($\pm{0.12}$) & 44.50($\pm{0.07}$) & 46.83($\pm{0.06}$)\\
     & SERank-b & \textbf{45.14}($\pm 0.13$) & \textbf{45.60}($\pm 0.11$) & \textbf{47.80}($\pm 0.09$) \\
    \hline
     \multirow{6}*{Web10K} & LambdaMART (LightGBM) & \textbf{46.20}($\pm 0.13)$) & \textbf{46.23}($\pm 0.09)$) & \textbf{48.33}($\pm 0.11)$) \\
     & RankNet & 39.71($\pm 0.09$) & 40.94($\pm 0.07$) & 43.37($\pm 0.05)$)\\
     & GSF(1) & 41.52($\pm 0.12$) & 42.12($\pm 0.11$) & 44.79($\pm 0.09$)\\
     & GSF(64) & 42.75($\pm 0.14$) & 42.65($\pm 0.12$) & 44.87($\pm 0.13$)\\
     \cline{2-5}
     & SERank & 40.39($\pm 0.13$) & 41.94($\pm 0.11$) & 44.46($\pm 0.09$)\\
     & SERank-b & \textbf{43.09}($\pm 0.14$) & \textbf{43.35}($\pm 0.12$) & \textbf{45.77}($\pm 0.10$) \\
  \bottomrule
\end{tabular}
\end{table*}

\indent These two loss functions are used in the experimentation section, and we leave analysis with other loss functions on our work to future study.

\section{Experiments} \label{experiment}
In this section, we first outline the datasets, baseline models and hyperparameters used in our experiments. We then compare our proposed model with baseline models on model effectiveness and complexity. Finally, we try to explain why our method works.

\subsection{Datasets}
The first dataset used in our experiments is MSLR-Web30K \cite{qin2013introducing}, which is a publicly available learning-to-rank dataset that contains more than 30000 queries. The relevance labels take 5 values from 0 (irrelevant) to 4 (perfectly relevant). We discard queries with no relevant documents. There are 136 dense features per query-document pair. The number of documents within one query is variable, which is on average 120. During training, we limit at most 200 documents per query, but when evaluate we use all. In this dataset, there are 5 folds containing the same data, and each fold randomly splits to train, validation and test set. Following \cite{ai2019learning} we use Fold1 in our experiments since results from other folds are similar. MSLR-Web10K \cite{qin2013introducing} is another learning-to-rank dataset that is similar to MSLR-Web30K but contains fewer samples (10,000 queries). And we will also report the experimental results on this dataset.  \newline
\indent The second dataset, Zhihu dataset, is a real-world learning-to-rank dataset created from Zhihu (a Chinese Question Answering community, www.zhihu.com) search log. We randomly sample user click logs from a week's traffic, and create a dataset with 2 million queries which is 60x larger than Web30K. For each query, there are on average 16 documents. The relevance label is obtained from the user’s click which is 1 when one document is clicked otherwise 0. Different from the Web30K dataset which mainly consists of dense features, the Zhihu dataset contains 129 dense features as well as sparse id features, like query and document title token ids. The dataset is randomly splited into three sets: train, validation and test set. The validation set and test set contain 5$\%$ of queries respectively. We report metrics in NDCG@1,5,10 \cite{jarvelin2002cumulated} on test set for all experiments.

\subsection{Models}
We have compared our method with multiple existing learning-to-rank models including tree-based methods and DNN models. The tree-based model used in our experiments is implemented in lightGBM \cite{ke2017lightgbm}. RankNet \cite{burges2010ranknet} and GSF \cite{ai2019learning} with different group size are used as DNN baseline methods. \newline
\indent For Web30K and Web10K dataset, the hyperparameters we used in LambdaMART (LightGBM) are consistent with previous work  \cite{bruch2019revisiting}. We train at most 1000 trees, then select the best number of trees by NDCG@5 on the validation set. For Zhihu dataset, we tune hyperparameters including learning rate, max number of leaves per tree and max number of trees to select the best model based on NDCG@5 of the validation set. \newline
\indent We implement RankNet and GSF models by TF Ranking \cite{pasumarthi2019tfranking}. RankNet is a multi-layer feed-forward fully-connected DNN model with Pairwise Logistic Loss (equation \ref{equation:pairwise-loss}). For the Web30K and Web10K dataset, we implement GSFs with the same hyperparameters as previous work \cite{ai2019learning}, which is a three layers (64,32,16) DNN model with batch normalization. For the Zhihu dataset, we use 7 dense layers (layer dims from 1024 to 16) with ReLU activation without batch normalization but we normalize the raw input features before feeding them into model. We transform sparse id features into fixed embedding and then apply average pooling on the sparse id emebdding features, then we concatenate dense input layer and embedding layer to obtain the final input layer. GSF models with various group sizes share the same model hyperparameters with RankNet except for the loss function which is softmax loss for Web30K dataset and $\lambda$-weighted Pairwise Logistic Loss for Zhihu dataset. For SERank we also use softmax loss on the Web30K dataset and $\lambda$-weighted Pairwise Logistic Loss for Zhihu dataset respectively. \newline
\indent For our proposed SERank model shown in Figure \ref{fig:senet-a}, based on the multi-layer feed-forward fully-connected structure, we further add the SE block followed by every dense layers. The shrinkage rate is 2 by default. The batch size is set to 128 and Adagrad \cite{fazayeli2014adaptive} optimizer is chosen with a learning rate of 0.5 for all DNN models. For the Web30K dataset, we train 30000 steps and select the best model on NDCG@5 of validation sets and then report predicted results on test set. And for the Zhihu dataset, we train 10 epoches for every model and evaluate on the test set using the last checkpoint.

\subsection{Comparison with Baseline Models}

In Table \ref{tab:web30k}, we compare our proposed method with other existing methods including tree-based models and DNN models on the Web30K and Web10K dataset. For GSF(1) and GSF(64), we cite the metrics reported in \cite{ai2019learning}. In addition, we further fine-tuned these two models and report our results which show 95$\%$ bootstrapped confidence intervals. For Web30K dataset, our proposed method (SERank-b) achieved the best result of DNN models which significantly outperform RankNet and GSF(1) by 3.5$\%$ and 1.9$\%$ on NDCG@5 (measured from 1 to 100) respectively, measured by paired t-test with p-value threshold of 0.05. And it also slightly surpass GSF(64) by 0.28$\%$. Because of the advantage of dense features and rather small data volume, the tree-based model LambdaMART (LightGBM) performs best on Web30K. This phenomenon is also observed in the related works \cite{ai2019learning}\cite{pasumarthi2019attentive}, where the evaluation score of their methods are lower than LambdaMART on Web30K dataset but performs opposite on the industrial dataset which is much larger than Web30K. Besides this, we can observe that our improved version SERank-b (with SE-b block) outperforms the original SERank (with SE block) by a large margin. In the following part, we adopt the improved SERank-b by default. On the Web10K dataset, the results show a similar tendency as Web30K which indicating the strong robustness of our proposed approach. \newline
\begin{table}
  \caption{\small{Comparison of test NDCG with baseline models on Zhihu dataset. For GSF and SERank models, we use pairwise logistic loss with $\lambda$-weight to train the model.}}
  \small
  \label{tab:zhihu}
  \begin{tabular}{l|ccc}
    \toprule
    Model & NDCG@1 & NDCG@5 & NDCG@10\\
    \midrule
    LambdaMART (LightGBM) & 51.16 & 60.89 & 66.63 \\
    GSF(1) & 53.23 & 63.28 & 69.24 \\
    GSF(2) & 53.15 & 63.21 & 69.10 \\
    GSF(64) & 51.91 & 62.35 & 68.39 \\
    SERank-b & \textbf{53.41} & \textbf{63.40} & \textbf{69.33} \\
  \bottomrule
\end{tabular}
\end{table}
\indent In Table \ref{tab:zhihu}, we compare the performance of SERank with other methods on the Zhihu dataset.  Similar to the scenario in Airbnb\cite{haldar2019applyingairbnb} and Gmail Search\cite{ai2019learning}, we also experiment SERank in the industrial dataset,  which is 60 times larger than Web30K. This magnitude of dataset makes the DNN model more advantageous than the tree-based LambdaMART. 
As the result shows, our proposed method outperforms all other methods. In terms of the results of GSF, we try different group sizes for GSF and find that as group size increases the performance becomes even worse and these similar phenomenons are also reported in \cite{pasumarthi2019self}.

\subsection{Model Efficiency}
Approaches that try to take advantage of cross-document interactions usually lead to huge increasing on computation cost, which is sensitive in online serving. We use FLOPs as the computation cost metric to evaluate model efficiency between SERank and other baseline models. The result of FLOPs is related to the input shape and model complexity. Therefore, we compute FLOPs in a forward pass of a  document sequence for one query in Web30K dataset, which has an input shape of [document\_size, feature\_size], where document size is 200 and feature\_size is 136 here. In Table \ref{tab:flops}, the FLOPs and test set NDCG@5 on the Web30K in dataset. Compared with GSF(1), the GSF(64) and the SERank improve relative 3.68$\%$ and 4.32$\%$ on NDCG@5 respectively, but GSF(64) cost 45.01 times FLOPs while SERank is only 1.75 times. In other words, the increasing computational cost of SERank is little over GSF(1) compared to the GSF(64) model, while the performance improvement is more significant. Combined with Figure \ref{fig:flops}, we can draw the conclusion that our proposed model is more effective and efficient on leveraging the Sequencewise information. \newline
\begin{table}
  \caption{\small{Comparison of normalized GFLOPs and rank metrics between SERank and GSFs on Web30K dataset}}
  \small
  \label{tab:flops}
  \begin{tabular}{l|cc}
    \toprule
    Model & $\Delta$ NDCG@5 & $\Delta$ FLOPs \\
    \midrule
    GSF(1) & - & - \\
    GSF(64) & 3.68$\%$ & 45.01 times \\
    SERank-b & 4.32$\%$ & 1.75 times \\
  \bottomrule
\end{tabular}
\end{table}

\subsection{Model Stability}
\begin{table}
  \caption{\small{Comparison of rank metrics between scoring with full docs and scoring with remaining docs}}
  \small
  \label{tab:stability}
  \begin{tabular}{l|ccc}
    \toprule
    Model & NDCG@1 & NDCG@5 & NDCG@10\\
    \midrule
    SERank-b-base & 44.85($\pm{0.06}$) & 47.83($\pm{0.04}$) & 51.53($\pm{0.05}$) \\
    SERank-b-remain-doc  & 45.01($\pm{0.05}$) & 47.84($\pm{0.03}$) & 51.48($\pm{0.03}$) \\
  \bottomrule
\end{tabular}
\end{table}
For ranking models utilizing cross-document information, it is important to study their ranking stability, i.e, the remaining documents should be ranked as the same order if some documents are removed from the candidate set. To examine the SERank's stability, we randomly mask out 50$\%$ of documents for each query in Web30K's test set and predict scores of the remaining ones. The NDCG metric is denoted as SERank-b-remain-doc in Table \ref{tab:stability}. On the contrary, we compare NDCG metric of the remained documents with their scores under no-missing circumstance (denoted as SERank-b-base). As Table \ref{tab:stability} shows, the NDCG is very close between these two conditions, which demonstrates that missing documents have little impact on the ranking order of other documents. Since this experiment has verified the robustness of our proposed work, we leave the stability test of GSF in future work.

\subsection{Ablation Study} \label{ablation}
In the previous section, we draw the conclusion that our proposed SERank is both effective and efficient. And we believe that the SE block which consists of squeeze structure and excitation structure plays the key role. In this section, we conduct an ablation study on the Web30K dataset to gain a better understanding of their roles.

\begin{table}
  \caption{\small{Effect of squeeze and excitation on Web30K dataset}}
  \small
  \label{tab:seblock}
  \begin{tabular}{l|ccc}
    \toprule
    Model & NDCG@1 & NDCG@5 & NDCG@10\\
    \midrule
    SERank-b & \textbf{45.14} & \textbf{45.60} & \textbf{47.81} \\
    SERank-b-W/O-Squeeze & 42.84 & 44.02 & 46.45 \\
    SERank-b-W/O-Excitation & 44.87 & 44.9 & 47.2 \\
  \bottomrule
\end{tabular}
\end{table}

\subsubsection{Effect of Squeeze}
\begin{figure}
    \centering
        \begin{subfigure}[t]{0.225\textwidth}
        \includegraphics[width=\textwidth]{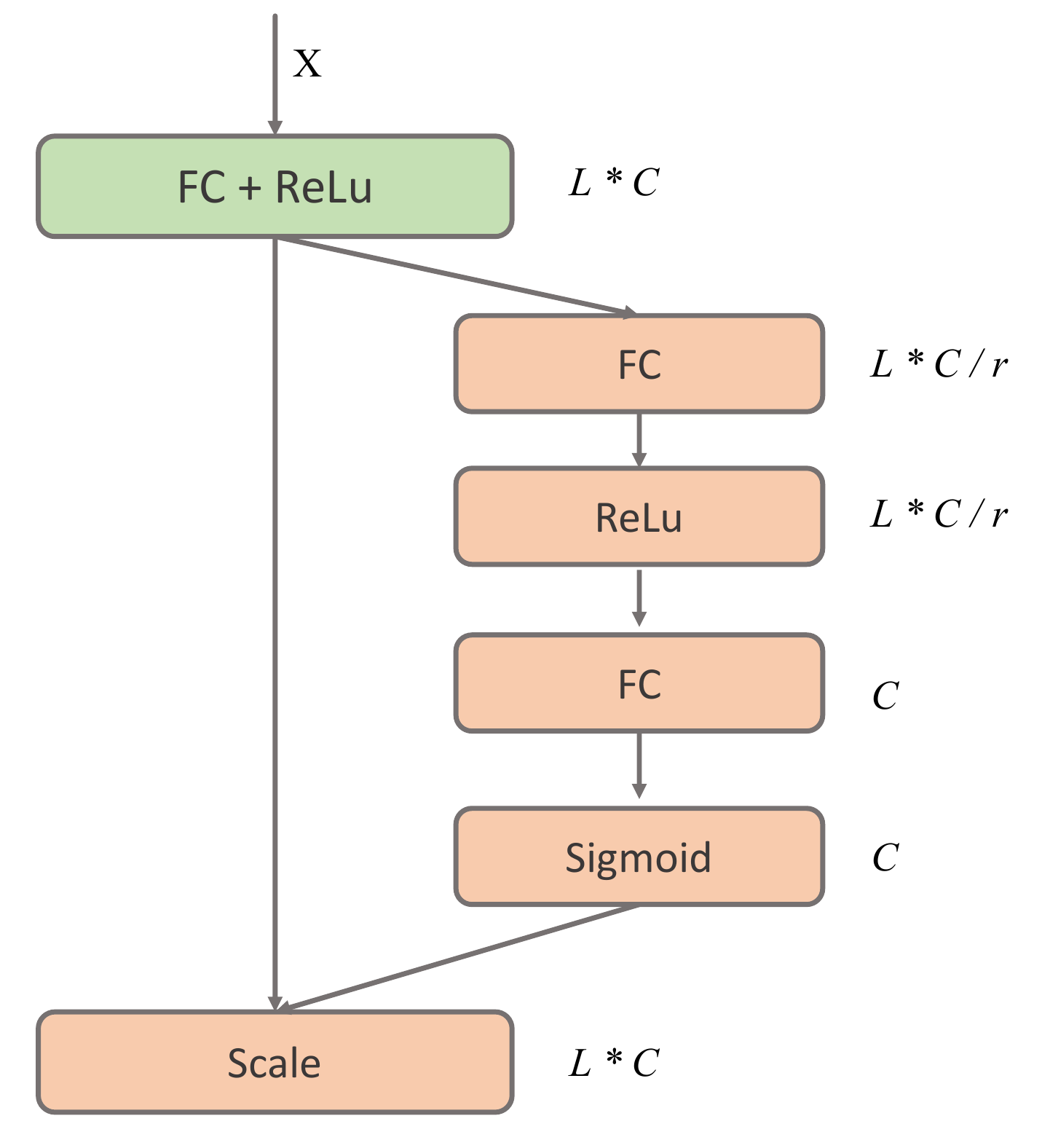}
        \caption{\small{SE block without squeeze}}
        \label{fig:nosqueeze}
    \end{subfigure}
    \begin{subfigure}[t]{0.2\textwidth}
        \centering
        \includegraphics[width=\textwidth]{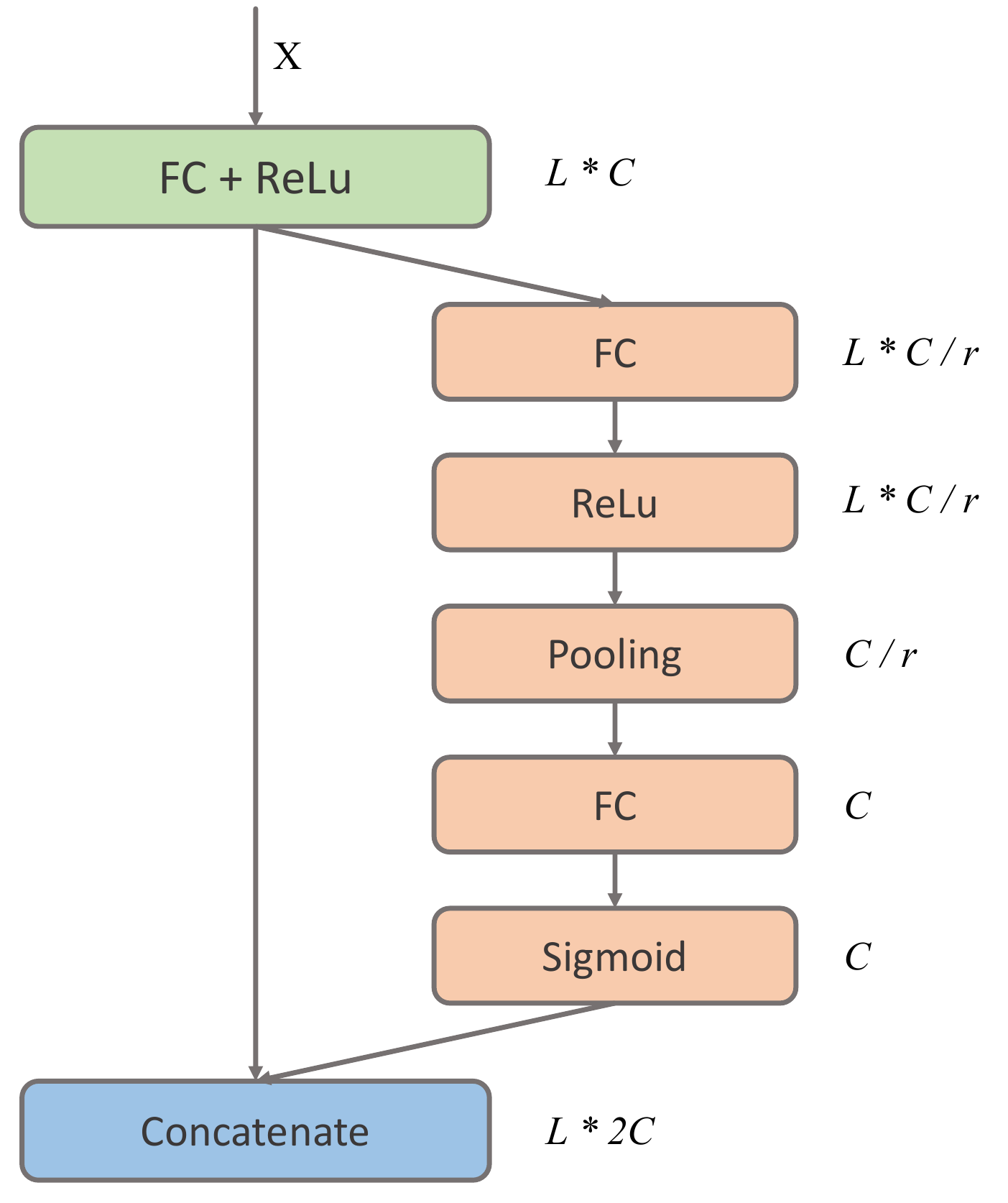}
        \caption{\small{SE block without excitation}}
        \label{fig:noexcitation}
    \end{subfigure}
    \caption{\small{Ablation study on Squeeze-And-Excitation operation}}
\end{figure}

Squeeze structure is a key component of the SE block. By pooling operation, the SE block can capture Sequencewise information to figure out the feature importance under current context. To verify our hypothesis, we experiment with a variant of the SE block which has no squeeze operation as Figure \ref{fig:nosqueeze} shows. Unlike the standard version, because the pooling operation is removed, the excitation outputs of the variant version are independent for different documents thus one document can not gain information from the others. As shown in Table \ref{tab:seblock}, the NDCG@5 of the SERank model without squeeze operation is significantly lower than the standard one, which proves the effectiveness of the squeeze operation.

\subsubsection{Effect of Excitation}
The excitation operation helps model learn the feature importance by applying the sigmoid function on the output of the dense layer and multiply with the original input of the SE block. By replacing the multiply operation with concatenate operation (concatenate the sigmoid out with the SE block’s inputs) as shown in Figure \ref{fig:noexcitation}, we can study the importance of excitation structure. Although this variant model can also benefit from Sequencewise information, it loses the physical meaning that the sigmoid outputs of excitation are the signal of the feature importance. In Table \ref{tab:seblock} we compare the ranking metrics of the SERank-b-W/O-Excitation with the standard one. As we can see, the performance of the SERank-b-W/O-Excitation model deteriorates, which is good proof of the effectiveness of excitation operation. 

\subsection{Online A/B Testing}
\begin{table}
    \small
    \centering
    \caption{\small{Online clicked search ratios at position 1,3,5}}
    \begin{tabular}{l|ccc}
        \toprule
        Metric    & Click@1  & Click@3  & Click@5 \\
        \midrule
        Increase       &  0.95\% & 0.4\% & 0.2\% \\ 
        P-value       &  0.001 & 0.002 & 0.004  \\ 
    \bottomrule
    \end{tabular}
    \label{tab:ab}
\end{table}
Besides evaluating SERank on the benchmark datasets, we further validate its performance by deploying it at the search engine of Zhihu, which is one of the largest Question Answering communities in China. We train SERank as well as the basic DNN model with nearly 20 million queries collected from one day's real search traffic. The baseline DNN model has three hidden layers with \{128,64,32\} hidden units with ReLU activation functions respectively, and neither dropout nor batch normalization were taken into training in both SERank and baseline DNN model. To eliminate position bias from click data, we adopt the method proposed by Zhao et al. \cite{zhao2019recommending}, which de-bias click data by adding an independent position aware tower in the model.\newline 
\indent We compared results of SERank with baseline model in term of the clicked search ratio at different positions, which is the percentage of clicked search sessions at top 1,3,5 among all search sessions. The higher of the clicked search ratio means the model performs better. To make online A/B testing reliable, we use p-value to test the significance of the two ranking models. Finally, to remove randomness between different traffic and ensure the results are comparable between traffic groups, we set two groups of study and two groups of control with equal amount of traffic. \newline 
\indent As shown in Table \ref{tab:ab}, the SERank model significantly outperform baseline DNN model at clicked search ratios. The clicked search ratios at position 1,3,5 are enhanced by relatively 0.9\%, 0.4\% and 0.2\% respectively. The p-value of all the metrics are smaller than 0.01 which indicates obvious significance of SERank. Therefore, our proposed SERank indeed improved ranking quality over the baseline DNN model.

\section{Conclusions} \label{conclusion}
In this paper, we propose a learning-to-rank approach denoted as SERank which aims at leveraging Sequencewise information to enhance the ranking metrics. We conduct a series of experiments comparing our proposed model with the existing methods on multiple datasets. The experimental results show that the SERank, which is not only effective but also efficient, outperforms the existing method in a statistically significant manner both on publicly available datasets and the large scale real-world dataset. Finally, the online A/B test shows that the SERank can significantly improve user experience in a large-scale industrial search engine.

\bibliography{sample-base}
\bibliographystyle{ACM-Reference-Format}
\end{document}